\documentclass[aps,prl,reprint,superscriptaddress]{revtex4-2}
\usepackage[T1]{fontenc}
\usepackage{mathptmx}

\usepackage{graphicx}

\usepackage[version=4]{mhchem}
\usepackage{textcomp}
\usepackage{dcolumn}
\usepackage{bm}
\usepackage{esdiff}
\usepackage{natbib}
\usepackage{booktabs}
\usepackage[colorlinks=true,citecolor=blue,linkcolor=blue,urlcolor=blue,pdfhighlight =/O]{hyperref}
\begin{document}
\title{Identifying $s$-wave pairing symmetry in single-layer FeSe from topologically trivial edge states}

\author{Zhongxu Wei}
\altaffiliation{These authors contributed equally to this work.}
\affiliation{Department of Physics, Southern University of Science and Technology, Shenzhen 518055, China}
\affiliation{State Key Laboratory of Low-Dimensional Quantum Physics, Department of Physics, Tsinghua University, Beijing 10008, China}

\author{Shengshan Qin}
\altaffiliation{These authors contributed equally to this work.}
\affiliation{Kavli Institute of Theoretical Sciences, University of Chinese Academy of Sciences, Beijing 100049, China}
\affiliation{CAS Center for Excellence in Topological Quantum Computation, University of Chinese Academy of Sciences, Beijing 100049, China}

\author{Cui Ding}
\affiliation{Beijing Academy of Quantum Information Sciences, Beijing 100193, China}
\affiliation{State Key Laboratory of Low-Dimensional Quantum Physics, Department of Physics, Tsinghua University, Beijing 10008, China}

\author{Jiangping Hu}
\affiliation{Beijing National Research Center for Condensed Matter Physics, and Institute of Physics, Chinese Academy of Sciences, Beijing 100190, China}
\affiliation{Kavli Institute of Theoretical Sciences, University of Chinese Academy of Sciences, Beijing 100049, China}
\affiliation{CAS Center for Excellence in Topological Quantum Computation, University of Chinese Academy of Sciences, Beijing 100049, China}

\author{Yujie Sun}\email{sunyj@sustech.edu.cn}
\affiliation{Department of Physics, Southern University of Science and Technology, Shenzhen 518055, China}

\author{Lili Wang}\email{liliwang@mail.tsinghua.edu.cn}
\affiliation{State Key Laboratory of Low-Dimensional Quantum Physics, Department of Physics, Tsinghua University, Beijing 10008, China}

\author{Qi-Kun Xue}\email{xueqk@sustech.edu.cn}
\affiliation{Department of Physics, Southern University of Science and Technology, Shenzhen 518055, China}
\affiliation{State Key Laboratory of Low-Dimensional Quantum Physics, Department of Physics, Tsinghua University, Beijing 10008, China}
\affiliation{Beijing Academy of Quantum Information Sciences, Beijing 100193, China}

\begin{abstract}

Determining the pairing symmetry of single-layer FeSe on SrTiO$_3$ is the key to understanding the enhanced pairing mechanism; furthermore, it guides exploring new superconductors with high transition temperatures ($T_\mathrm{c}$). Despite significant efforts, it remains controversial whether the symmetry is the sign-preserving $s$- or the sign-changed $s_{\pm}$-wave. Here, we investigate the pairing symmetry of single-layer FeSe from a topological point of view. Using low-temperature scanning tunneling microscopy/spectroscopy, we have systematically characterized the superconducting states at isolated edges and corners of single-layer FeSe. The tunneling spectra collected at edges and corners exhibit full gap and substantial dip, respectively, demonstrating the absence of topologically non-trivial edge/corner modes. According to the theoretical calculation, these spectroscopic features are strong evidence for the sign-preserving $s$-wave pairing.

\end{abstract}

\maketitle

\section{Introduction}
\begin{figure}[ht!]
  \centering
  \includegraphics[width=\linewidth]{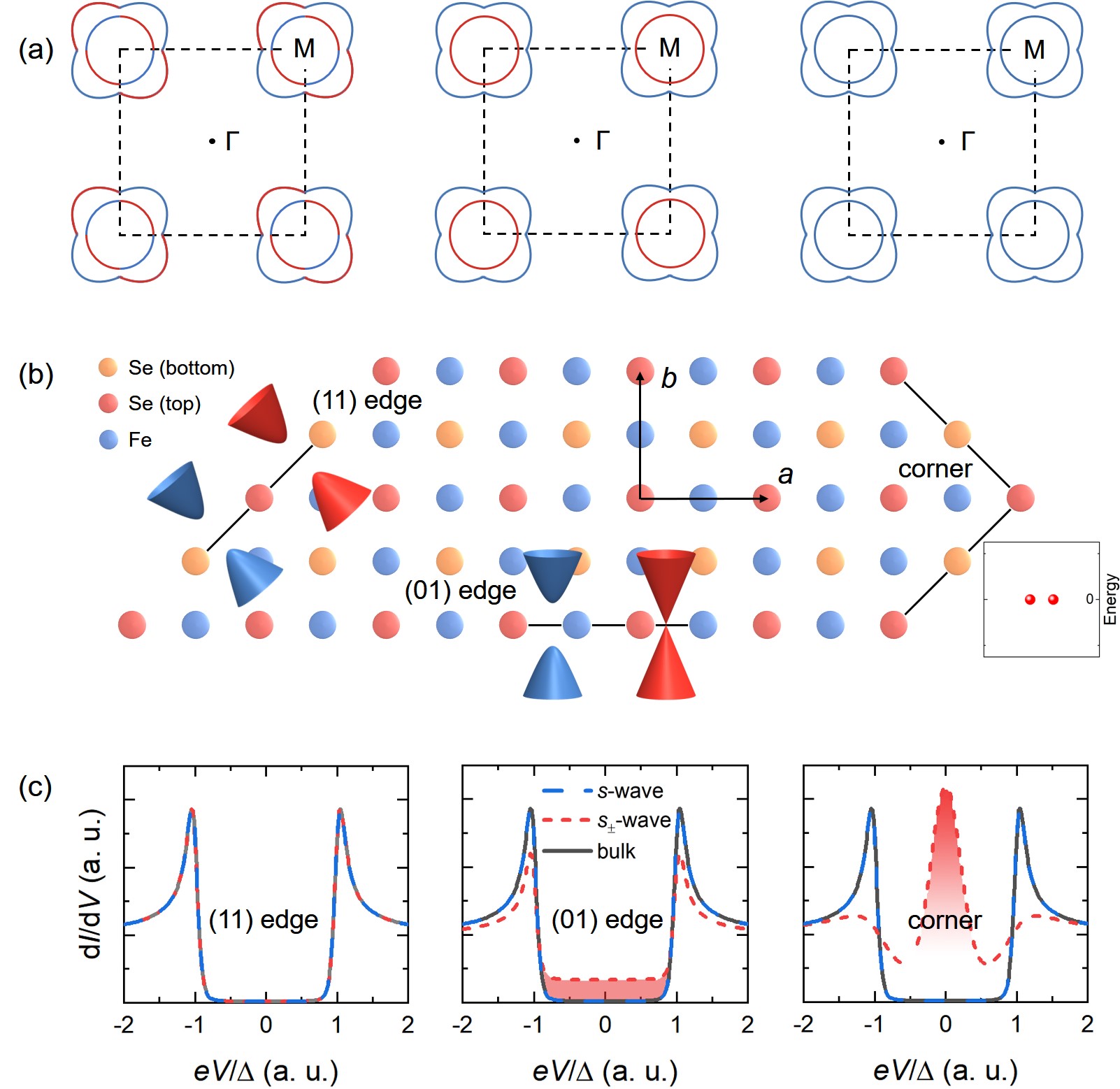}
  \caption{(a) Schematic illustration of nodeless $d$-, sign-changed $s_{\pm}$- and sign-preserving $s$-wave pairings with band hybridization taken into account. Different colors denote the opposite sign of order parameters. (b) Definition of edges and the corner. The blue and red cones indicate the band structures at various edges in the case of $s$- and $s_{\pm}$-wave pairing, respectively. The lower right panel indicates zero-energy modes for the $s_{\pm}$-wave case. (c) Supposed tunneling spectra at the (11) and (01) edges and the corner for the $s$- (blue) and $s_{\pm}$-wave (red) cases. The grey lines are spectra away from the edge/corner.}
  \label{fig1}
  \end{figure}

Heavily electron-doped iron chalcogenides A$_x$Fe$_2$Se$_2$ \cite{Guo2010S,Ying2011S} (A = alkali metal), (Li$_{1-x}$Fe$_x$)OHFeSe \cite{Dong2015L,Huang2017S}, and single-layer FeSe on SrTiO$_3$ \cite{Wang2012I,Zhang2014I} have attracted significant attention due to their high $T_\mathrm{c}$ and simple Fermi surface that consisted only of electron pockets centered at the M point of the Brillouin zone (BZ) \cite{Zhang2011N,He2013P,Niu2015S}. Identifying the pairing symmetry of these materials is crucial to reveal the pairing mechanism fully. Previous angle-resolved photoemission spectroscopy \cite{Zhang2011N,He2013P,Niu2015S} and scanning tunneling microscopy/spectroscopy (STM/S) measurements \cite{Li2012K,Du2016S,Gong2019O} have consistently demonstrated anisotropic gap without nodes, suggesting nodeless $d$-, sign-changed $s_{\pm}$- and sign-preserving $s$-wave states as competitive candidates [Figure \ref{fig1}(a)]. The nodeless $d$-wave state is unlikely because the strength of spin-orbit coupling (SOC) between electron pockets is comparable to that of pairing \cite{Mazin2011S,Borisenko2016D,Agterberg2017R}. Moreover, the behavior of Caroli-de Gennes-Matricon states in the vortex core disfavors $d$-wave pairing \cite{Chen2020O,Zhang2021O}. On the other hand, the controversy over the two leading contenders, \emph{i.e.}, sign-preserving $s$- and sign-changed $s_{\pm}$-wave, remains unresolved, despite extensive impurity-scattering investigations \cite{Fan2015P,Yan2016S,Du2018S,Liu2018E,Liu2019S,Liu2019D,Zhang2020S} including measurements employing the recently developed quasiparticle interference technique \cite{Fan2015P,Yan2016S,Chen2018R,Du2018S,Gu2018S,Liu2020H}. The conflict situation stems from the difficulties in proving the practical magnetic or nonmagnetic nature of the impurities under investigation \cite{Liu2018E} and distinguishing bound states under weak scattering potential \cite{Chen2016E}. In addition, although phase-sensitive tests based on tunneling junctions have been proposed much earlier \cite{Golubov2013D}, hitherto no experimental result has been reported for iron chalcogenides. Overall, the pairing nature of these materials is still an open issue, and more deterministic characterization is desired.

Recently, we have developed a promising way from a topological point of view to settle the long-lasting debate between the $s$- and $s_{\pm}$-wave pairing \cite{Qin2022T}. Specifically, as shown in Figure \ref{fig1}(b), the inversion center in Se-Fe-Se triple-layer is at the nearest Fe-Fe bond center rather than the Fe site, which naturally generates the Rashba-type SOC between the next-nearest-neighbor Fe sites \cite{Zhang2014H,Wu2017D,Agterberg2017R,Zhang2020Sy,Qin2022T}. Accompanied by the unique lattice structure and SOC, anomalous band degeneracies along the BZ boundary develop. As a consequence, a sign-changed $s_{\pm}$-wave pairing leads to a second-order superconducting state which supports two Dirac cones at the (01) edge and a pair of Majorana zero-energy modes at the corner between the (11) and (1$\overline{1}$) edges [Figure \ref{fig1}(b)] \cite{Qin2022T}. In contrast, sign-preserving $s$-wave states remain topologically trivial even in the presence of inversion symmetric Rashba SOC. Therefore, the gapless edge modes and zero-energy corner modes, which can be directly probed by STM/S [Figure \ref{fig1}(c)], serve as the smoking-gun evidence to distinguish the $s$- and $s_{\pm}$-wave pairing.

In this work, we investigate the superconducting states at isolated edges and corners of single-layer FeSe. Our spectroscopic investigations demonstrate that the superconductivity gets suppressed with moving close to the isolated edges but remains fully gapped along with both (01) and (11) edges. Furthermore, there is no evidence for corner Majorana modes at the intersection between the (11) and (1$\overline{1}$) edges. Based on our recent theoretical calculation \cite{Qin2022T}, these topologically trivial superconducting states support the sign-preserving $s$-wave pairing in single-layer FeSe.

\section{Methods}
High-quality FeSe thin films were grown on Nb-doped SrTiO$_3$(001) (0.05 wt. \%) substrates by molecular beam epitaxy. The surface is TiO$_2$ terminated after heating to 1150 {\textcelsius} for 15 minutes. Then FeSe films were fabricated by co-evaporating Fe (99.995\%) and Se (99.999\%) from standard Knudsen cells onto SrTiO$_3$ kept at 450 {\textcelsius}. The growth rate was $\sim$ 0.024 unit cell (uc) per minute. By controlling the coverage to be less than 1 uc, well-defined edges with various orientations form between FeSe and vacuum. After annealing at 480 - 490 {\textcelsius} for 4.5 hours, the sample was transferred \emph{in situ} to the STM chamber. Topographic images were obtained by the constant-current mode. The tunneling spectra were measured using a standard lock-in technique with a sample bias ($V_\mathrm{s}$) of 50 mV, a tunneling current ($I_\mathrm{t}$) of 500 pA, and a bias modulation of 0.5 mV at 937.2 Hz. The differential conductance d$I$/d$V$($V$) of all spectra is calibrated proportionally by scaling d$I$/d$V$($V_\mathrm{s}$) to $I_\mathrm{t}$/$V_\mathrm{s}$. Commercial Pt/Ir tips were calibrated on Ag films before performing STM/S experiments. All STM/S experiments were carried out at 4.8 K.

\section{Experimental Results}

\begin{figure}[ht!]
	\centering
	\includegraphics[width=\linewidth]{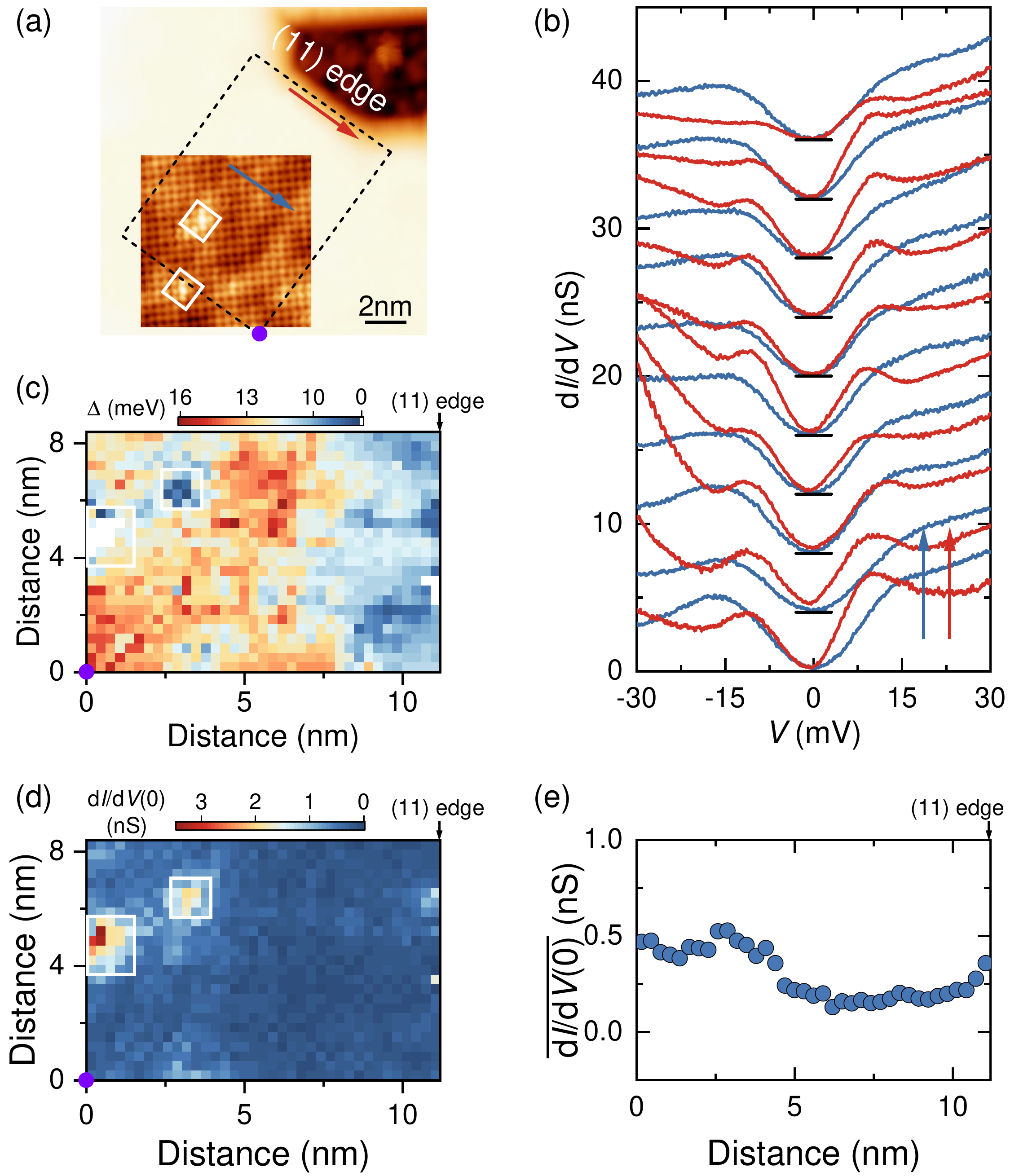}
	\caption{(a) STM topographic image ($V_\mathrm{s} = 1$ V, $I_\mathrm{t} = 50$ pA) of the (11) edge. Inset: atomically resolved image ($V_\mathrm{s} = 50$ mV, $I_\mathrm{t} = 500$ pA) of FeSe. (b) Two sets of tunneling spectra measured along the blue and red arrows in (a). The black lines indicate individual zero conductance. (c)-(d) Energy gap map (c) and zero-bias conductance map (d) obtained from the spectroscopic mapping over the area outlined by the black dashed box in (a). The purple circles mark the origin point. (e) Column-averaged zero-bias conductance as a function of the distance to the (11) edge.}
	\label{fig2}
\end{figure}

Figure \ref{fig2}(a) shows a typical topographic image of single-layer FeSe. The dark area in the upper right corner is the exposed SrTiO$_3$ substrate. The apparent height of the step between FeSe and SrTiO$_3$ is $\sim$ 660 pm at a $V_\mathrm{s}$ of 1 V, which is slightly larger than the out-of-plane lattice constant of FeSe (550 pm) due to the varied density of states \cite{Yu2021S}. This edge is along the (11) direction, as judged from the atomically resolved Se lattice structure shown in the inset image. Figure \ref{fig2}(b) shows two sets of tunneling spectra far from (blue) and near (red) the edge. There is no significant difference in the zero-bias conductance of these tunnel spectra. We also record tunneling spectra over the area indicated by the black dashed box in Figure \ref{fig2}(a). Figure \ref{fig2}(c) and Figure S1 in Supplemental Material \cite{supplemental} display the position-dependent superconducting energy gap ($\Delta$), where $\Delta$ is extracted by half the distance between coherence peaks and Dynes formula \cite{Dynes1984T}, respectively. Obviously, $\Delta$'s near the (11) edge are smaller than those in bulk, which may be due to the lattice discontinuity. Nevertheless, the zero-bias conductance, our main concern, is uniform in real space  [Figure \ref{fig2}(d)]. Excluding anomalies induced by local defects marked by white boxes [Figures \ref{fig2}(a), (c) and (d)], the column-averaged zero-bias conductance is almost position-independent [Figure \ref{fig2}(e)]. The absence of finite constant conductance across the superconducting gap meets the expectation [Figure \ref{fig1}(c)], providing a reference for analyzing the related results at the (01) edge.

\begin{figure}[ht!]
	\centering
	\includegraphics[width=\linewidth]{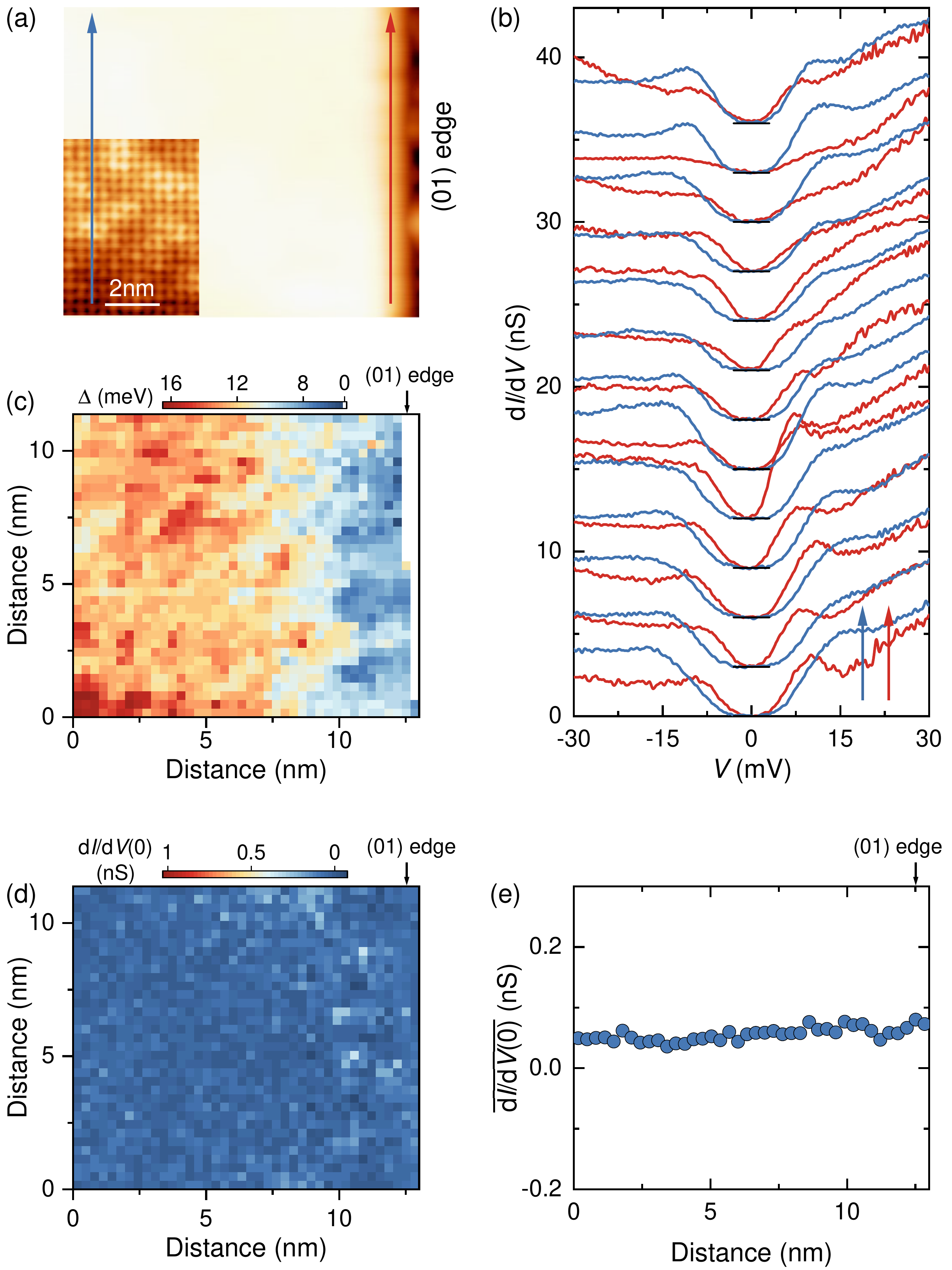}
	\caption{(a) STM topographic image ($V_\mathrm{s} = 1$ V, $I_\mathrm{t} = 50$ pA) of the (01) edge. Inset: atomically resolved image ($V_\mathrm{s} = 50$ mV, $I_\mathrm{t} = 500$ pA) of FeSe. (b) Two sets of tunneling spectra measured along the blue and red arrows in (a). (c)-(d) Energy gap map (c) and zero-bias conductance map (d) obtained from the spectroscopic mapping over the same field of view as in (a). (e) Column-averaged zero-bias conductance as a function of the distance to the (01) edge.}
	\label{fig3}
\end{figure}

Figure \ref{fig3}(a) depicts an STM image of (01) edge. This specular edge extends more than 12 nm, providing an excellent opportunity to study the edge modes. As shown in Figure \ref{fig3}(b), the coherence peak shrinks near the (01) edge while the zero-bias conductance remains unchanged. For further verification, we collect a spectroscopy map over the same field of view as shown in Figure \ref{fig3}(a). The evolutions of $\Delta$ derived from coherence peaks spacing and Dynes formula are presented in Figure \ref{fig3}(c) and Figure S1 in Supplemental Material \cite{supplemental}, respectively. Similar to the results obtained near the (11) edge, $\Delta$ decreases when approaching the (01) edge. To capture the trace of the edge modes, we check the shape of tunneling spectra and find that almost all spectra near the (01) edge remain fully gapped. Figure \ref{fig3}(d) shows the zero-bias conductance map, and Figure \ref{fig3}(e) presents the column-averaged values. It is obvious that zero-bias conductance is position-independent. The difference in values of zero-bias conductance in Figures \ref{fig2}(e) and \ref{fig3}(e) could be due to the different sample batches and tip apexes. We also collect the spectra at other (01) edges and get the same results repeatedly (see section B of Supplemental Material \cite{supplemental}). Therefore, the superconducting states at the (01) edge are topologically trivial.

\begin{figure}[ht!]
	\centering
	\includegraphics[width=\linewidth]{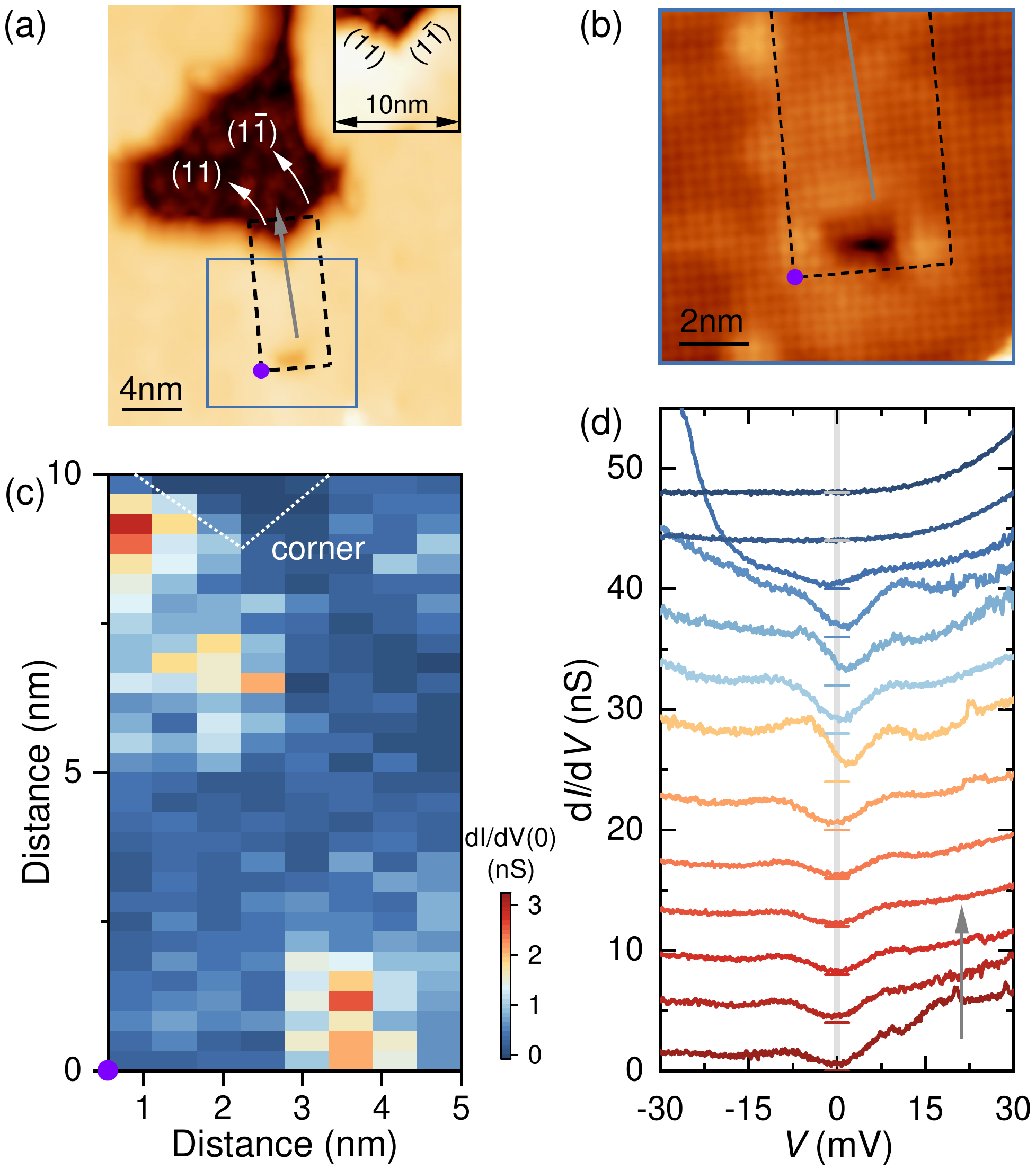}
	\caption{(a) STM topographic image ($V_\mathrm{s} = 1$ V, $I_\mathrm{t} = 50$ pA) of the corner. Inset: zoom-in image. (b) Atomically resolved image ($V_\mathrm{s} = 50$ mV, $I_\mathrm{t} = 500$ pA) taken from the area outlined by the blue box in (a). (c) Zero-bias conductance map obtained from the spectroscopic mapping over the area indicated by the black dashed box in (a). The purple circles mark the origin point. (d) The tunneling spectra collected along the grey arrow in (a).}
	\label{fig4}
\end{figure}

To further confirm the topological properties of single-layer FeSe, the corner states are studied. Figure \ref{fig4}(a) shows a topographic image containing a corner formed by the (11) and (1$\overline{1}$) edges, which can be verified by the Se lattice shown in Figure \ref{fig4}(b). Due to the epitaxial growth of FeSe, intersected (11) and (1$\overline{1}$) edges are rarely observed and generally extend only a few nanometers. Figure \ref{fig4}(d) presents a set of tunneling spectra measured along the grey arrow in Figure \ref{fig4}(a), with the top two spectra taken on the SrTiO$_3$(001) surface. The superconductivity is gradually suppressed when approaching the corner, but without anomalies around 0 mV, that is, zero-bias conductance peak (ZBCP) as supposed for $s_{\pm}$-pairing is missing. For careful verification, we perform differential conductance mapping at 0 mV over intersections between the (11) and (1$\overline{1}$) edges. As exemplified in Figure \ref{fig4}(c), the differential conductance map corresponding to the area outlined by the black box in Figure \ref{fig4}(a) shows that the conductance is relatively uniform. We emphasize that the results are highly reproducible (see section C of Supplemental Material \cite{supplemental}), confirming the topologically trivial superconductivity in single-layer FeSe.

\section{Discussion}

We first clarify the effects of the substrate on the detection of the edge/corner modes. The coupling between SrTiO$_3$ and FeSe may lead to the non-observation of topological modes in two ways: (i) The edge/corner states leak into the substrate. (ii) The non-uniform strain arising from structural instability of SrTiO$_3$ \cite{Erdman2002T,Tao2016N} introduces local mirror symmetry breaking, which is detrimental to the topological superconductivity. Case (i) is unlikely because the SrTiO$_3$(001) surface contributes no electronic states near the Fermi energy [Figure \ref{fig4}(d) and Figures S2-S3 in Supplemental Material \cite{supplemental}], which indicates that the edge/corner modes will exponentially decay in the substrate and have to be localized around the edge/corner. For case (ii), the local mirror symmetry breaking, if any, is relatively weak as confirmed by topographic images [Figures \ref{fig2}(a), \ref{fig3}(a), \ref{fig4}(a) and Figures S2-S5 in Supplemental Material \cite{supplemental}]. In addition, even if the two Dirac edge modes at the (01) edge could hybridize and be gapped out under weak mirror symmetry breaking, the single Majorana Kramers' pair located at the isolated corner can exist stably \cite{Geier2018S}. On the other hand, the spectra collected under $I_\mathrm{t} / V_\mathrm{s} = 10$ nS with our equipment can capture anomalies in differential conductance of meV-scaled energy [Figure \ref{fig2}(d) and Ref. \cite{Liu2018E,Liu2020Z,Cheng2020A}]. Therefore, the spectra measured at the edge/corner intrinsically reflect the topological property of single-layer FeSe.

The topological property of single-layer FeSe provides essential information on its pairing symmetry \cite{Qin2022T}. In the case of $s_{\pm}$-wave pairing, the second-order topological superconductivity arises in centrosymmetric single-layer FeSe with the help of additional glide-plane and mirror symmetries \cite{Qin2022T}. Specifically, two Dirac cones and one single Majorana Kramers' pair are expected respectively at the (01) edge and the corner between the (11) and (1$\overline{1}$) edges. The Dirac cones contribute finite energy-independent density of states within the superconducting gap, and the Majorana Kramers' pair leads to a quantized ZBCP in the tunneling spectrum. In the case of $s$-wave pairing, however, the superconducting states at all edges and corners are topologically trivial, \emph{i.e.}, fully gapped. Therefore, as summarized in Table \ref{tab1}, studying the spectroscopic features at edges/corners is a practical way to distinguish between the sign-preserving $s$- and sign-changed $s_{\pm}$-wave states. The tunneling spectra shown in Figures \ref{fig2} $\sim$ \ref{fig4} definitively exclude the existence of edge modes and Majorana modes. Consequently, we unambiguously conclude that the pairing symmetry of single-layer FeSe is the sign-preserving $s$-wave rather than the sign-changed $s_{\pm}$-wave.

\begin{table}[!htbp]
    \caption{Features of tunneling spectra of single-layer FeSe at various edges and corners.}
    \label{tab1}
    \centering
    \begin{tabular}{ c c c c }
        \toprule[1pt]
          & ~~~~(11) edge~~~~ & ~~~~(01) edge~~~~ & ~~~~(11) \& (1$\overline{1}$) corner~~~~ \\
        \midrule[1pt]
        $s$-wave & gapped & gapped & gapped \\
        \midrule[0.5pt]
        $s_{\pm}$-wave & gapped & gapless & ZBCP \\
        \bottomrule[0.5pt]
    \end{tabular}
\end{table}

It is worthy to note that previous works, Refs. \cite{Chen2020O} and \cite{Ge2019E}, have also investigated edge states of single-layer FeSe. Ref. \cite{Chen2020O} reports the fully gapped spectra along two kinds of (01) edges. The first kind of edge is formed by the 1 uc and 2 uc FeSe, and the second kind of edge consists of 1 uc FeSe situated on either side of the SrTiO$_3$ step. In the former case, the superconducting bottom layer of the 2 uc FeSe side \cite{Wang2017S} smoothly extends to the 1 uc FeSe side, indicating that such configuration is actually the edge of the non-superconducting upper layer of the 2uc FeSe side. In the latter case, FeSe on adjacent SrTiO$_3$ terraces is non-separate, since the thickness of FeSe (550 pm) is larger than the step height of SrTiO$_3$ (390 pm). The physics near such edge is elusive, and whether there is a response in electronic states to the topological superconductivity needs more detailed study. Under a simplest assumption that FeSe films on both sides are weakly linked, the tunneling spectra at the edge are expected to be fully gapped, regardless of the pairing symmetry. In contrast, all edges and corners investigated in our work are constructed of 1uc FeSe and vacuum, which is an ideal condition for detecting the edge/corner modes. Ref. \cite{Ge2019E} reports a pair of emergent conductance peaks located near the superconducting gap at the (01) edge. However, the conductance peak, which can only be resolved after a normalization done by subtracting the spectrum far from the edge, is most likely due to the reduction in superconducting gap near the edge and therefore are not related to topological properties (see section D of Supplemental Material \cite{supplemental}).

\section{Conclusion}
In conclusion, we fabricate high-quality single-layer FeSe by molecular beam epitaxy and investigate the electronic structures at different edges and corners by STM/S. We neither observe gapless edge modes at the (01) edges nor detect ZBCP at the corners between the (11) and (1$\overline{1}$) edges. The topologically trivial superconducting states are solid evidence supporting the sign-preserving $s$-wave pairing symmetry of single-layer FeSe. More delicate experiments are to be designed to identify the pairing glue supporting $s$-wave state such as orbital fluctuation, phonon, etc. \cite{Chen2015E,Yamakawa2017S,Yamakawa2020D}. Finally, our achievements also pave a promising way to determine the pairing symmetry of other iron-based superconductors such as single-layer Fe(Se,Te) \cite{Li2015I}, K$_x$Fe$_2$Se$_2$ \cite{Tang2015S} and (Li$_{1-x}$Fe$_x$)OHFeSe.

\section{Acknowledgements}
This work is supported by the National Natural Science Foundation of China (Grants No. 12074210, 51788104, 12141402 and No. 11790311) and the National Basic Research Program of China (Grant No. 2017YFA0303303).

\bibliography{ref} 

\end{document}